\documentclass[11pt]{article}

\usepackage{graphicx}
\usepackage{epstopdf}
\usepackage{amsmath}
\usepackage{amssymb}
\usepackage{amsfonts}
\usepackage{amsthm}
\usepackage[usenames]{color}
\usepackage{array}

\usepackage[
colorlinks=true,
linkcolor=blue,
urlcolor=blue,
filecolor=blue,
citecolor=red,
pdfstartview=FitV,
pdftitle={},
pdfauthor={},
pdfsubject={},
pdfkeywords={},
pdfpagemode=None,
bookmarksopen=true
]{hyperref}

\usepackage{epsfig}

\usepackage{hyperref}

\newcommand{\bea}{\begin{eqnarray}}
\newcommand{\eea}{\end{eqnarray}}
\newcommand{\ba}{\begin{array}}
	\newcommand{\ea}{\end{array}}
\newcommand{\ee}{\end{equation}}
\numberwithin{equation}{section}

\begin{document}

\begin{flushright}
	\texttt{\today}
\end{flushright}

\begin{centering}
	
	\vspace{2cm}
	
	\textbf{\Large{
%			  Information paradox and its resolution 
Islands in Flat-Space Cosmology }}
	
	\vspace{0.8cm}
	
	{\large   Sanam Azarnia$^{a}$, Reza Fareghbal$^{a,b}$, Ali Naseh$^{b}$, Hamed Zolfi$^{b}$ }
	
	\vspace{0.5cm}
	
	\begin{minipage}{.9\textwidth}\small
		\begin{center}
			
			{\it  $^{a}$Department of Physics, 
				Shahid Beheshti University, 1983969411,  
				 Tehran , Iran \\
			
			$^{b}$School of Particles and Accelerators, Institute for Research in Fundamental Sciences (IPM)
			P.O. Box 19395-5531, Tehran, Iran}
			
			\vspace{0.5cm}
			{\tt  sanam.azarnia@gmail.com, r$\_$fareghbal@sbu.ac.ir, naseh,\hspace{-1.5mm} hamedzolphy@ipm.ir}
			\\ 
			
		\end{center}
	\end{minipage}

	%\end{center}

\begin{abstract}
Flat-space cosmologies  (FSCs) are solutions to  three-dimensional theories of gravity  without  cosmological constants that have  cosmological horizons. A detector located near the timelike singularity of the spacetime can absorb Hawking modes that are created near the horizon. Continuation of this process will eventually cause the entropy of the radiation to be larger than the entropy of the FSC, which leads to the information paradox.
In this paper, we resolve this paradox for the FSC using the island proposal. To do this, we couple an auxiliary  flat bath system to this spacetime in timelike  singularity so that Hawking modes are allowed to enter the bath  and the entropy of radiation can be measured in its asymptotic region where gravity is also weak. We show that adding island regions that receive the  partners of Hawking modes cause the  entropy of radiation to follow  a Page curve which leads to resolving the information paradox. Moreover, we design a quantum teleportation protocol by which one can extract the information residing in islands.  
\end{abstract}

\end{centering}

\newpage

\tableofcontents

\section{Introduction}
Recently, Einstein gravity without cosmological constants in three dimensions has been considered  with more interest. This theory does not have black hole solutions \cite{Ida:2000jh}, but in addition to Minkowski spacetime, other solutions can be found that have a horizon \cite{Cornalba:2002fi}. The event horizons for these asymptotically flat spacetimes  are cosmological horizons, hence they are called flat-space cosmologies (FSCs).
These spacetimes can be locally converted to Minkowski spacetime, and their relationship to Minkowski spacetime is similar to what BTZ black holes have to the anti-de Sitter (AdS) spacetimes. In fact, the FSCs are orbifolds of  Minkowski spacetime and can be converted to it with a shift and boosts  in coordinates \cite{Cornalba:2002nv}.
Moreover, the FSCs can be obtained from the BTZ black hole by taking the flat-space limit (zero cosmological constant limit or infinite AdS radius limit), which makes a large number of its properties easily proven from the asymptomatic AdS case using the flat-space limit.\\

One of the reasons for paying attention to  the  FSC spacetimes  is that they appear in the holography of asymptotically flat spacetimes. It has been recently  proposed that asymptotically flat spacetimes in three dimensions are dual to the states of a two-dimensional field theory with an   ultra  relativistic symmetry \cite{Bagchi:2010zz},\cite{Bagchi:2012cy}.
In fact, the origin of this proposal  is  that asymptotic symmetries  at null infinity of asymptotically flat spacetimes   in three dimensions are given by the Bondi-Metzner-Sachs group \cite{Ashtekar:1996cd},\cite{Barnich:2006av}. This group is isomorphic  to  another group in two  dimensions, which is obtained by Inonu-Wigner contraction of the conformal group \cite{Leblond65}. This contraction is done by taking the zero limit of the light speed. Hence this group   is ultrarelativistic, and  is called the Carrollian conformal group \cite{Duval:2014uoa}. We call the field theory with this symmetry Carrollian conformal field theory and the corresponding holography flat/Carrollian conformal field theory. Various aspects of this holography have been explored so far, a list of which can be found in the references of a recent article \cite{Bagchi:2021qfe}.\\ 

It is possible to define entropy and temperature for the cosmological horizon of FSC spacetime. The entropy in this case is similar to black holes and is  proportional  to the area of the horizon. In Ref. \cite{Bagchi:2012xr}, it was  shown that this entropy can be obtained by counting the number of corresponding states in the dual field theory. The FSCs, like black holes, can have Hawking radiation \cite{Cornalba:2002fi}. The causal structure of this spacetime (up to asymptotic regions) and the type of its horizons are exactly the same as de Sitter spacetimes. The study of radiation from the cosmological horizon of de Sitter spacetimes has been done for the first time in the  paper \cite{Gibbons:1977mu}. This study can be  generalized to the FSCs, which ultimately leads to the definition of temperature for the cosmological horizon that is proportional  to its surface gravity.\\

Due to the timelike singularity of FSC spacetime, the Hawking mode created near the horizon is reflected from it and after a while reaches its partner inside the horizon. The result of this meeting is the purification of their state and therefore there is no notion of information paradox. This picture is similar to the one for the large black holes in AdS spacetimes where the Hawking mode is reflected from the asymptotic timelike boundary and the total state after a while becomes a thermofield double state at fixed temperature. The eternal AdS black holes become radiative by coupling their asymptotic boundary to an auxiliary flat spacetime (bath system). In this coupled system, the Hawking mode instead of reflection from the asymptotic boundary is absorbed in the bath and therefore the entanglement entropy of radiation becomes larger and larger, same as the one for an evaporating black hole. This is an information paradox since unitarity limits the
maximum entropy of a black hole to be the Bekenstein-Hawking entropy.\\

Interestingly, this paradox can be resolved with the help of the island proposal \cite{Penington:2019npb,Almheiri:2019psf}. The main idea in this proposal is the existence of new regions (islands) containing the partner of Hawking mode, which could be added to the entanglement wedge of radiation collected in the bath. Since, by this addition, the state of Hawking mode and its partner is purified to a Bell state, the increasing of entanglement entropy of radiation is stopped and information paradox is resolved. The boundary of these new regions are located either inside or outside the horizon and are called quantum extremal surfaces (QES) since they are minimum of the following generalized entropy functional for the radiation \cite{Penington:2019npb,Almheiri:2019psf},
\bea\label{island}
S_{\text{Rad}} = \text{Min}\hspace{.2mm} \bigg\{\text{Ext}\bigg[\frac{\text{Area}(\partial I)}{4G_{\text{N}}}+S_{\text{vN}}(R \cup I)\bigg]\hspace{.2mm}\bigg\}.
\eea
The Area refers to the area of codimension 2 boundary surface of island, $\partial I$, and $S_{\text{vN}}[R\cup I]$ is the von Neumann entropy of the quantum state of combined radiation and island systems computed in the effective semiclassical theory. The island here refers to any number of regions, including zero, contained in the gravitational region. The combination of an area term with quantum matter entanglement entropy is known as generalized entanglement entropy. The procedure for applying this formula is extremizing the right-hand side of (\ref{island}) with respect to the position of the boundary of island then minimizing over all extremal surfaces. It is worth mentioning that albeit this procedure comes from the AdS/CFT, i.e., Ryu-Takayanagi formula \cite{Ryu:2006bv} and its extensions \cite{Hubeny:2007xt, Engelhardt:2014gca}, but actually it can be applied for any QFT coupled to  gravity. Marvelously, this proposal has come out proud in all the examples that have been used, i.e., for two-dimensional black holes 
of JT (Jackiw-Teitelboim) gravity \cite{Almheiri:2019psf,Almheiri:2019hni,Almheiri:2019yqk,Bousso:2019ykv, Hollowood:2020cou}, two-dimensional asymptotically flat  solutions of the Callan–Giddings–Harvey–Strominger model \cite{Callan:1992rs,Gautason:2020tmk,Anegawa:2020ezn,Hartman:2020swn}, higher dimensional spacetimes in Einstein gravity \cite{Almheiri:2019psy,Hashimoto:2020cas},  higher curvature gravities \cite{Alishahiha:2020qza}, massive gravities \cite{Nam:2021bml}, and de Sitter horizons \cite{Balasubramanian:2020xqf,Hartman:2020khs, Geng:2021wcq,Aalsma:2021bit,Kames-King:2021etp} \footnote{The cosmological circuit complexity in the presence of islands is studied in \cite{Choudhury:2020hil}. Also, the role played by mutual information of subsystems on the Page curve is explored in \cite{Saha:2021ohr}.}. For all of them, the entanglement entropy of radiation follows the Page curve \cite{Page:1993wv} and therefore the unitarity is restored. Albeit the whole procedure is done in the semiclassical limit, since the information paradox is resolved by a coherent connection between gravity and quantum mechanics, it is very important to the understanding of quantum gravity.\\

In this paper we would like to extend the previous studies on the island formula to the FSC spacetimes. For this purpose, we make its timelike singularity transparent by coupling it to an  auxiliary bath and then examine the island proposal. We will show that the entanglement entropy of radiation follows the Page curve thanks to the appearance of a new quantum extremal surface outside of the horizon. We will discuss that to have an agreement with quantum focusing conjecture, the boundary of the island cannot appear inside the horizon in the coupled system and we also design a quantum teleportation protocol by which one can extract the information residing in the island.  
%The extra minus sign in first law of FSC thermodynamics, (\ref{first law CH}), is crucial to have them all. 
We organize the paper as follows: In Sec. \ref{FSCspacetime}, we explain more carefully the FSC geometry and its thermodynamics characteristics such as temperature, Bekenstein-Hawking entropy, and also the associated first law. Also, the proper coordinates that can be analytically continued to the whole geometry of two-sided eternal FSC geometry coupled to the bath will be introduced. Moreover, the procedure of emitting the Hawking radiation from the cosmological horizon will be explained. In Sec. \ref{islandsection}, we apply the island formula and find the new quantum extremal surface outside the horizon. After calculating the Page time and scrambling time we also design a procedure to extract information from the island. The last section, \ref{Discussion}, is devoted to the conclusion and future directions.
\section{Three-dimensional FSC}\label{FSCspacetime}
\subsection{The FSC solution and its conformal structure}
The gravitational theory that we consider in this paper is Einstein gravity without cosmological constant in three dimensions. 
In addition to the Minkowski spacetime, which is an obvious solution, it is easy to check that the following metric also satisfies its equations of motion \cite{Cornalba:2002fi}: 
\begin{equation}\label{metric of FSC}
ds^2 =-{dr^2\over f(r )} + f(r )dt^2+ r^2\big(d\phi-N_\phi(r) dt\big)^2,
\end{equation}
with
\begin{equation}
f(r)=\dfrac{\hat r_+^2(r^2-r_0^2)}{r^2},\qquad N_\phi(r)=\dfrac{r_0\hat r_+}{r^2},
\end{equation}
and $r_0$ and $\hat r_+$ are two constants. The function $f(r)$ vanishes at $r=r_0$ and for $r>r_0$ the coordinate $r$ is timelike while  $t$ is spacelike. Thus $r=r_0$ is a cosmological horizon  and the geometry given by \eqref{metric of FSC} is known as FSC. There is also an intrinsic singularity at $r=0$, which is timelike. The two parameters $r_0$  and $\hat{r}_+$ are related to the mass $\mathcal{M}$ and angular momentum $\mathcal{J}$ of the FSC by
\begin{equation}
\hat r_+=\sqrt{8G_{N}\mathcal{M}},\qquad r_0=\sqrt{\dfrac{2G_{N}}{\mathcal{M}}}|\mathcal{J}|. 
\end{equation}

\begin{figure}[h]
\centering
\includegraphics[scale=.6]{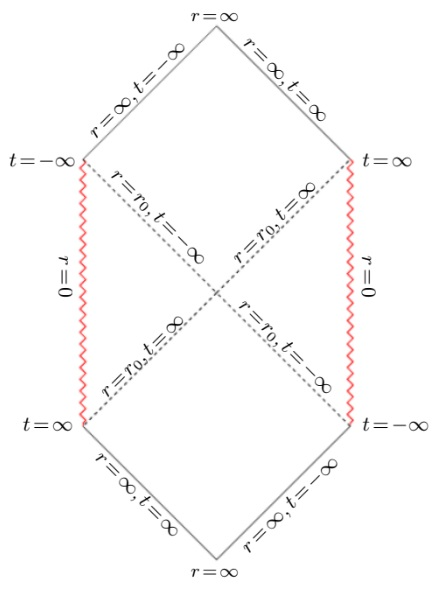} 
\caption{Penrose diagram of FSC solution. The dashed lines indicate the cosmological horizon. The solid black lines and wiggly red lines are corresponded to the past-future null infinities and timelike singularities, respectively.}\label{FSCwithoutbath}
\end{figure}

It is possible to find a locally well-defined transformation which maps FSC \eqref{metric of FSC} to the Minkowski spacetime. This transformation is a combination of shift and boost of the coordinates and shows that FSC only encompasses part of the Minkowski geometry. Accordingly, FSC is known as the shift-boost orbifold of the Minkowski spacetime. Moreover, FSC is given by taking the flat-space limit from the BTZ black hole,
\begin{equation}\label{BTZ metric}
ds^2=-\dfrac{(r^2-r_+^2)(r^2-r_-^2) dt^2}{\ell^2 r^2}+\dfrac{\ell^2 r^2 dr^2}{(r^2-r_+^2)(r^2-r_-^2)}+r^2\left(d\phi-\dfrac{r_+r_-}{\ell r^2} dt\right)^2,
\end{equation}
where $r_+$ and $r_-$ are the radii of outer and inner horizons and $\ell=-4/\Lambda$ is the AdS radius according to the negative cosmological constant $\Lambda$ . The radii $r_\pm$ are given in terms of mass $M$ and angular momentum $J$ of BTZ as follows
\begin{equation}\label{radii BTZ in M J}
M=\dfrac{r_+^2+r_-^2}{8G_{N}\ell^2},\qquad J=\dfrac{r_+\hspace{.5mm}r_-}{4G_{N}\ell}.
\end{equation}
Plugging  \eqref{radii BTZ in M J} in \eqref{BTZ metric} and taking the $\ell\to\infty$  limit results in \eqref{metric of FSC}. It is clear from \eqref{radii BTZ in M J} that at large AdS radius, $r_-\to r_0$ and $r_+\to\ell \hat r_+ $. Thus the region between outer horizon and AdS boundary is vanished by taking the flat-space limit. Furthermore, it is worth noting that the $\phi$-constant surfaces  in geometry \eqref{metric of FSC} are spacelike for $r<r_0$ and timelike for $r>r_0$. In order to have a well-defined spacelike periodic coordinate in all parts of spacetime, we define a new coordinate $\psi$,
\begin{equation}\label{psi}
\psi=\phi-\dfrac{\hat r_+}{r_0}t,
\end{equation}
which by that the metric of $\psi$-constant two-dimensional surfaces becomes
\bea\label{newFSC}
ds^2 = \frac{\hat{r}_+^2 (r^2-r_0^2)}{r_0^2} \hspace{.5mm}dt^2 -\frac{r^2 }{\hat{r}_+^{2}(r^2-r_0^2)}\hspace{.5mm}dr^2.
\eea 
The Penrose diagram of maximally extended FSC spacetime, (\ref{metric of FSC}), is depicted in Fig. \ref{FSCwithoutbath}. The corresponding Kruskal coordinate $U$ and $V$ for the upper wedge is given by\footnote{This coordinates are written by taking the flat space limit from the Kruskal coordinates of BTZ black hole introduced in \cite{Carlip:1995qv}.}
\bea\label{kruskalfsc1}
U=e^{\frac{\hat r_+^2}{r_0}\hspace{.2mm}u(t,r)},\qquad V=e^{\frac{\hat r_+^2}{r_0}\hspace{.2mm}v(t,r)},
\eea
where
\bea\nonumber
u(t,r)=\dfrac{1}{\hat r_+^2}\left(r+\dfrac{r_0}{2}\log{\dfrac{r-r_0}{r+r_0}}\right)-t,\hspace{1cm}
v(t,r)=\dfrac{1}{\hat r_+^2}\left(r+\dfrac{r_0}{2}\log{\dfrac{r-r_0}{r+r_0}}\right)+t.
\eea
The definition of  $U$ and $V$ coordinates in  other  wedges of Fig. \ref{FSCwithoutbath} can be written by analytic continuation of \eqref{kruskalfsc1} as follows
\bea\label{Kruskall}
&&{U}_{\text{right}}(t,r) = {U}_{\text{top}}\left(t+i\pi\frac{r_0}{\hat r_+^2},r\right),\hspace{.85cm}{V}_{\text{right}}(t,r) = {V}_{\text{top}}(t,r),
\cr \nonumber\\
&&{U}_{\text{left}}(t,r) = {U}_{\text{top}}(t,r),\hspace{2.3cm}{V}_{\text{left}}(t,r) = {V}_{\text{top}}\left(t+i\pi\frac{r_0}{\hat r_+^2},r\right),
\cr \nonumber\\
&&{U}_{\text{bottom}}(t,r) = {U}_{\text{top}}\left(t+i\pi\frac{r_0}{\hat r_+^2},r\right),\hspace{.7cm}{V}_{\text{bottom}}(t,r) = {V}_{\text{top}}\left(t+i\pi\frac{r_0}{\hat r_+^2},r\right).
\eea
The temperature and entropy for the cosmological horizon $r_0$, are given by
\begin{equation}\label{temp and entropy of FSC}
T=\dfrac{\kappa}{2\pi}=\dfrac{\hat r_+^2}{2\pi r_0},\hspace{1cm} S_{\text{th}}=\dfrac{A}{4G}=\dfrac{\pi r_0}{2 G_{N}}.
\end{equation}
Interesting point is that both of the temperature and entropy of the cosmological horizon are given by taking the flat-space limit from the temperature and entropy of the BTZ inner horizon. This relation between the BTZ inner horizon and the cosmological horizon of FSC also has an impact on the first law of cosmological horizon. Similar to the inner horizon of BTZ \cite{Detournay:2012ug}, we can write
\begin{equation}\label{first law CH}
dM=-TdS+\Omega dJ,
\end{equation}
where $\Omega=\hat r _+/r_0$ is the angular velocity of the cosmological horizon.\vspace{.5cm}

\subsection{Radiation from the FSC horizon}    
\begin{figure}[h]
\centering
\includegraphics[scale=.6]{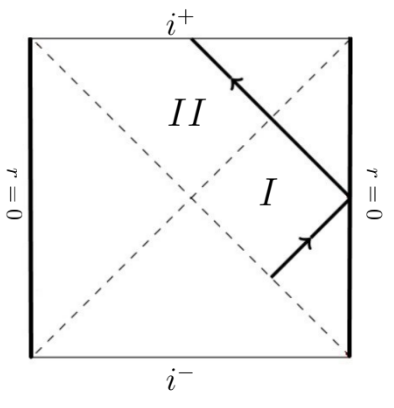} 
\caption{Penrose diagram of de Sitter spacetime}\label{ds}
\end{figure}

In order to study the Hawking radiation from the cosmological horizon of the FSC, we compare the causal structure of this spacetime, Fig.     \ref{FSCwithoutbath}, with the one for de Sitter (dS) spacetime, Fig. \ref{ds}. The metric of dS spacetime in the static coordinate is given by
\bea
ds^2=-\left(1-\dfrac{r^2}{\ell^2}\right)dt^2+\dfrac{dr^2}{\left(1-\dfrac{r^2}{\ell^2}\right)}+r^2 d\phi^2,
\eea
where $r=\ell$ locates its cosmological horizon. Thus in both of spacetimes in the region $I$ or  $r<r_{\text{CH}}$ (where $r_{\text{CH}}$ is the radius of cosmological horizon), $t$ is timelike, and $r$ is spacelike. However, dS has a spacelike boundary in region $ II $, at $r=\infty$. The radiation from cosmological horizon of dS spacetime has been studied in \cite{Gibbons:1977mu} by Gibbons and Hawking. The mechanism is very similar to the radiation from the event horizon of black holes, i.e., a pair of particle and antiparticle are created from the vacuum state near the horizon in region $ I $. Antiparticle with negative energy (with respect to the timelike Killing vector $\partial_t$) crosses the horizon and appears in region $ II $ where $\partial_t$ is now spacelike. Therefore, in region $II$ it can be considered as an ordinary particle. Moreover, the particle in region $I$ moves towards $r=0$ and can be detected  by a detector located near $r=0$. The flux of these detected particles is interpreted as the radiation from the cosmological horizon. The picture for the FSC solution is similar to \cite{Cornalba:2002fi} and one can deduce  Hawking radiation for its cosmological horizon with the temperature \eqref{temp and entropy of FSC}.
The absorption of radiation by the detector increases its entropy and  thus decreases  the entropy of the cosmological  horizon. This reduces the area of the horizon and according to the first law \eqref{first law CH} we can conclude that the mass of FSC increases.
%One way to explain this is that radiation can be thought of as the quantum tunneling of a particle from the outside of the cosmological horizon into the inside. It is clear that the particle moves inward can increase the mass of FSC. 
\section{Island and Resolving the Information Paradox}\label{islandsection}
\begin{figure}[h]
\centering\includegraphics[scale=.6]{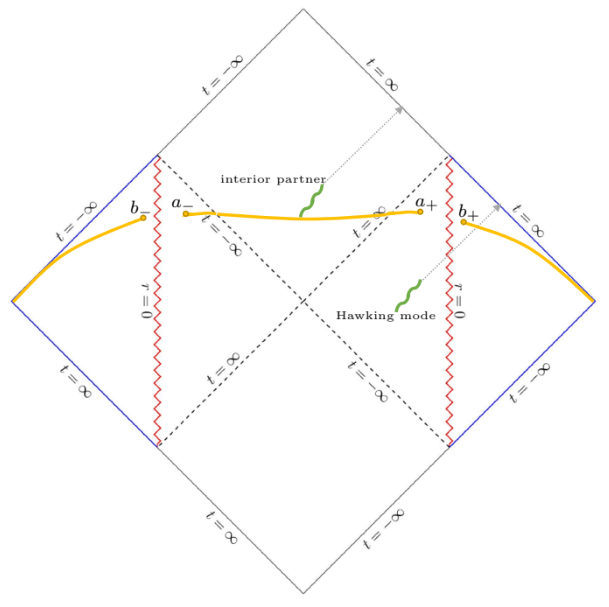} 
\caption{Penrose diagram of eternal FSC solution coupled to the bath. The Hawking mode and its interior partner are represented by green curved lines. The dashed line(s) and dotted line(s) indicate the location of horizon and the path of entangled modes, respectively.  The amount of energy emitted by cosmological horizon (Hawking radiation) is same as one falling in from the bath, therefore in all times the geometry is eternal solution.  Since moving the left side backwards in time and the right one forwards in time is an isometry, we can make left and right times equal by using this isometry.}\label{PenroseFSC}
\end{figure}

If there is no detector near the singularity, each Hawking mode created near the cosmological horizon at early times, will be reflected at the timelike singularity (wiggly red line in Fig.  \ref{FSCwithoutbath}) at later times. This implies a balance between emitting particles from the cosmological horizon and absorbing the ones reflected from the singularity. Accordingly, the total state is independent of the time. It is similar  to what happens for the eternal AdS black holes, where each Hawking mode is reflected from the timelike asymptotic boundary and then is absorbed by the black hole at a finite time. Similar to simulating the black hole evaporation in the eternal AdS black hole spacetime, we can connect a bath to the right and left singularities in a way that the common boundaries become transparent, see Fig.  \ref{PenroseFSC}. By ``bath" we mean a quantum mechanical system where we can neglect the gravitational effects. In this setup, the whole system starts in a Hartle-Hawking state (thermofield double state) and it is evolved forward in time on both sides. Now, the FSC is radiative. The Hawking radiation can be collected in the bath region and its full fine-grained entropy is given by the island formula (\ref{island}).
One technical issue in applying the formula (\ref{island}) is that in three-dimensional spacetimes, the entanglement entropy of quantum matter has an arealike UV divergence,
\bea
S_{\text{vN}} (R\cup I) = \frac{\text{Area}(\partial I)}{\varepsilon} + S_{\text{vN,fin}}(R\cup I),
\eea 
where $\varepsilon$ is the short distance cutoff scale. This divergence can be absorbed by renormalization of Newton constant, $1/4G_{\text{N}}+1/\varepsilon = 1/4G_{\text{N,ren}}$. Accordingly, the island formula (\ref{island}) changes to 
\bea\label{islandren}
S_{\text{Rad}} = \text{Min}\hspace{.2mm} \bigg\{\text{Ext}\bigg[\frac{\text{Area}(\partial I)}{4G_{\text{N,ren}}}+S_{\text{vN,fin}}(R \cup I)\bigg]\bigg\}.
 \eea 
Another technical issue is about the number of quantum matter (scalar) fields, $N \sim c$ where $c$ (some fixed positive constant for a unitary theory) denotes the central charge. In the following, we assume that $1 \ll N \ll S_{\text{th}}$, where in competition with the graviton the quantum matter has main contribution to the von Neumann entropy while at the same time the backreaction of matter fields on the geometry is negligible. Another technical issue is that the calculation of entanglement entropy of a three-dimensional quantum matter field for a region with several disconnected intervals is not an easy task. But here we are interested in the entanglement entropy of quantum fields on geometry with angular symmetry in the direction of the $\psi$ coordinate, (\ref{psi}). By choosing the expansion of quantum fields according to this symmetry, the reduced two-dimensional theory apart from the massless modes contains some massive Kaluza-Klein modes whose masses are given by angular momentum along the circle $S^{1}$. Since the entangling regions that are interesting for us here are far from each other, the contribution of those massive modes to entanglement entropy is negligible. Accordingly, we have a massless quantum field theory (CFT)\footnote{To be more precise, a specific type of 2D CFT, for example free fermions.} on two-dimensional geometry  (\ref{newFSC}) and therefore we can use the following two-dimensional formula \cite{Calabrese:2009ez} for the finite part of entanglement entropy of the union of regions $R_{+} =[b_{+},\infty]$, $R_{-}=[b_-,\infty]$ and  island $I=[a_-,a_+]$ in Fig.  \ref{PenroseFSC},
\bea\label{Svn}
S_{\text{vN,fin}} (R\cup I) = \frac{c}{6}\log \left(\frac{L(a_+,a_-)L(b_+,b_-)L(a_+,b_+)L(a_-,b_-)}{L(a_+,b_-)L(a_-,b_+)}\right),
\eea
where $L(p_1,p_2)$ denotes the proper geodesic length between two point $p_1(t,r)$ and $p_2(t,r)$. It is worth mentioning that to write the above expression we have used this fact that the whole system (FSC+bath) represents a pure state, actually the vacuum state. To have a vacuum state in the whole system, we work within a new null coordinates  $\mathbf{U}$ and $\mathbf{V}$. For  the top wedge in Fig.  \ref{FSCwithoutbath}, they are defined as
\bea\label{UV}
\mathbf{U}(t,r)=e^{\frac{2\pi}{\beta}\mathbf{u}(t,r)}= e^{-\frac{2\pi}{\beta}\left(t-r^{*}(r)\right)},\hspace{1cm} \mathbf{V}(t,r)=e^{\frac{2\pi}{\beta}\mathbf{v}(t,r)} = e^{\frac{2\pi}{\beta}\left(t+r^{*}(r)\right)},
\eea
where $\beta$ is the inverse of the temperature \eqref{temp and entropy of FSC} and
\bea\label{tortoise}
r^{*}_{\text{FSC}}(r) = \frac{\beta}{4\pi} \log \left(\mid\hspace{-1mm}r^2 -r_0^2\hspace{-1mm}\mid\right).
\eea
This definition  can also be analytically continued to other wedges in Fig.  \ref{FSCwithoutbath} by relations similar to \eqref{Kruskall}.
% for example, for the top-square (top-wedge) in figure \ref{PenroseFSC}, we have 
%\bea\label{Kruskal}
%U_{\text{top}}(t,r) \equiv e^{\frac{2\pi}{\beta}u(t,r)} =  e^{-\frac{2\pi}{\beta}\left(t-r^{*}(r)\right)},\hspace{2cm}V_{\text{top}}(t,r) \equiv e^{\frac{2\pi}{\beta}v(t,r)}= e^{\frac{2\pi}{\beta}\left(t+r^{*}(r)\right)},
%\eea
%where the corresponding two dimensional  metric (\ref{newFSC}) becomes
%\bea\label{Klike}
%ds^{2}_{\text{FSC}} = -\Omega^{-2} dU dV, \hspace{1cm}\Omega(r) = \frac{\hat{r}_{+}}{\sqrt{r^2-r_0^2}}\hspace{1mm}e^{\frac{2\pi r^{*}(r)}{\beta}}.
%\eea
The auxiliary bath system is in the thermal equilibrium with the FSC spacetime (\ref{newFSC}) and therefore for the whole system,  we can write the metric as 
\bea\label{bath}
ds^{2} = -\Omega^{-2}\hspace{1mm}d\mathbf{U} d\mathbf{V},
\eea
where\footnote{According to (\ref{omega}), the metric is not continuous at the singularity where we couple the two systems, but there the flux of energy is actually smooth by noting to the conformal symmetry of quantum matter field.}
\bea\label{omega}
&& 
\Omega_{\text{FSC}}(r) = \frac{\hat{r}_{+}}{\sqrt{r^2-r_0^2}}\hspace{1mm}e^{\frac{2\pi }{\beta}r_{\text{FSC}}^{*}(r)},\hspace{1.5cm}\Omega_{\text{Bath}} = \frac{2\pi}{\beta}\hspace{1mm}e^{\frac{2\pi }{\beta}r_{\text{Bath}}^{*}(r)},
\eea
and $r_{\text{Bath}}^{*}(r) = r$. Accordingly, the geodesic distance $L(p_1,p_2)$ in the whole system  becomes
\bea\label{L}
L(p_1,p_2) = \frac{\big(\mathbf{U}(p_2)-\mathbf{U}(p_1)\big)\big(\mathbf{V}(p_1)-\mathbf{V}(p_2)\big)}{\Omega(p_1)\hspace{1mm}\Omega(p_2)}.
\eea
The main reason to use the null coordinates  $\mathbf{U},\mathbf{V}$ instead of the null coordinates $U,V$ (\ref{kruskalfsc1}) is that for the latter one, the metric apart from the off-diagonal component also has another components. This implies that the geodesic length $L(p_1,p_2)$ would not have the simple form as (\ref{L}).    
 \subsection{Island outside the horizon of coupled system}\label{islandoutside}
Using the expression (\ref{L}) for the geodesic distance and for $b_+:(t_b,b),\hspace{1mm}a_+:(t_a,a),\hspace{1mm} b_-:(-t_b,b),\hspace{1mm} a_-:(-t_a,a)$ together with considering the possible quantum extremal surface in the right wedge \footnote{The proper coordinates are $\mathbf{U}_{\text{right}}$ and $\mathbf{V}_{\text{right}}$, according to  (\ref{Kruskall}) for analytical continuations.}, the finite part of the entanglement entropy (\ref{Svn}) becomes
\bea\label{SVN}
&&\hspace{-1cm}S_{\text{vN,fin}} = \frac{c}{6}\log\bigg[\frac{\beta^2 (r_0^2-a^2)}{4\pi^2\hat{r}_+^2}e^{-\frac{4\pi(t_a+t_b)}{\beta}}\left(1+e^{\frac{4\pi t_a}{\beta}}\right)^2\left(1+e^{\frac{4\pi t_b}{\beta}}\right)^2
\cr\nonumber\\
&&\hspace{.8cm}\times\hspace{1mm}\frac{\left(e^{\frac{2\pi}{\beta}\big(t_a+t_b+2b\big)}-e^{\frac{2\pi}{\beta}\big(2t_a+b+r^{*}(a)\big)}-e^{\frac{2\pi}{\beta}\big(2t_b+b+r^{*}(a)\big)}+e^{\frac{2\pi}{\beta}\big(t_a+t_b+2r^{*}(a)\big)}\right)^2}{\left(e^{\frac{2\pi}{\beta}\big(t_a+t_b+2b\big)}+e^{\frac{2\pi}{\beta}\big(b+r^{*}(a)\big)}+e^{\frac{2\pi}{\beta}\big(2t_a+2t_b+b+r^{*}(a)\big)}+e^{\frac{2\pi}{\beta}\big(t_a+t_b+2r^{*}(a)\big)}\right)^2}\hspace{.5mm}\bigg]\hspace{-1mm}.
\eea
Let us firstly explore the existence of quantum extremal surface at early times. At early times when $t_a/b,t_b/b \ll 1$, the von Neumann entropy (\ref{SVN}) simplifies to
\bea
S_\text{vN,fin} = \frac{c}{6}\log\bigg[\frac{4\beta^2 (r_0^2-a^2)}{\pi^2 \hat{r}_+^2}\frac{\left(e^{\frac{4\pi b}{\beta}}+e^{\frac{4\pi r^{*}(a)}{\beta}}-2e^{\frac{2\pi}{\beta}\big(b+r^{*}(a)\big)}\right)^2}{\left(e^{\frac{4\pi b}{\beta}}+2e^{\frac{2\pi}{\beta}\big(b+r^{*}(a)\big)}+e^{\frac{2\pi}{\beta}\big(t_a+t_b+2r^{*}(a)\big)}\right)^2}\bigg].
\eea
Therefore, the generalized entropy in these times reads
\bea
S_{\text{gen}} = \frac{\pi\hspace{1mm} a}{G_{\text{N,ren}}} + \frac{c}{6}\log\bigg[\frac{4\beta^2 (r_0^2-a^2)}{\pi^2 \hat{r}_+^2}\frac{\left(e^{\frac{4\pi b}{\beta}}+e^{\frac{4\pi r^{*}(a)}{\beta}}-2e^{\frac{2\pi}{\beta}\big(b+r^{*}(a)\big)}\right)^2}{\left(e^{\frac{4\pi b}{\beta}}+2e^{\frac{2\pi}{\beta}\big(b+r^{*}(a)\big)}+e^{\frac{2\pi}{\beta}\big(t_a+t_b+2r^{*}(a)\big)}\right)^2}\bigg],
\eea
that it might be extremized with respect to the location of quantum extremal surface, i.e., with respect to $a$ and $t_a$. The extremization with respect to $a$ gives
\bea\label{early1}
&&\frac{\partial S_{\text{gen}}}{\partial a} =\frac{1}{3(r_0^2-a^2)}\bigg(\frac{3\pi(r_0^2-a^2)-c\hspace{.5mm}G_{\text{N,ren}}\hspace{1mm} a}{G_{\text{N,ren}}}
\cr\nonumber\\
&&\hspace{1cm} +\frac{4c \left(e^{\frac{2\pi}{\beta}\big(3b+r^{*}(a)\big)}-e^{\frac{2\pi}{\beta}\big(b+3r^{*}(a)\big)}\right)a}{\left(e^{\frac{2\pi b}{\beta}}-e^{\frac{2\pi r^{*}(a)}{\beta}}\right)^2 \left(e^{\frac{4\pi b}{\beta}}+2e^{\frac{2\pi}{\beta}\big(b+r^{*}(a)\big)}+e^{\frac{2\pi}{\beta}\big(t_a+t_b+2r^{*}(a)\big)}\right)}\hspace{.5mm}\bigg) = 0,
\eea
and the extremization with respect to $t_a$ becomes
\bea\label{early2}
\frac{\partial S_{\text{gen}}}{\partial t_a} = -\frac{2\pi c\hspace{1mm}e^{\frac{2\pi}{\beta}\big(t_a+t_b+2r^{*}(a)\big)}}{3\beta \left(e^{\frac{4\pi b}{\beta}}+2e^{\frac{2\pi}{\beta}\big(b+r^{*}(a)\big)}+e^{\frac{2\pi}{\beta}\big(t_a+t_b+2r^{*}(a)\big)}\right)}=0.
\eea
From Eq. (\ref{early2}), it is clear that the only possibility for the boundary of island is $a = r_0$. But this value of $a$ apparently is not acceptable by the equation (\ref{early1}). This implies that there is no island at early times. In absence of the island, only the contribution of quantum matter to the generalized entanglement entropy (\ref{SVN}) remains
\bea\label{Sgentwosidewithout}
S_{\text{gen}} = S_{\text{vN,fin}} = \frac{c}{6}\log\left[\frac{\beta^2}{\pi^2}\cosh^{2}(\frac{2\pi t_b}{\beta})\hspace{.5mm}\right],
\eea
which at very early times  behaves as the following\footnote{The $\mathcal{O}(t^2)$ growth is similar to the time dependency of entanglement entropy after a global quench in very early times \cite{Cardy:2015}.}
\bea
S_{\text{gen}} \sim \frac{2\pi^2 c}{3}\hspace{1mm}\frac{t_b^2}{\beta^2},
\eea
while at late times (after a few thermal times) , entropy grows linearly 
\bea\label{linear}
S_{\text{gen}} \sim \frac{2\pi c}{3}\hspace{1mm}\frac{t_b}{\beta}. 
\eea
This linear growth at late times is nothing but the observation of Hawking to have  information paradox. To be more precise, because of this late time linear growth the finiteness of von Neumann entropy for a finite dimensional system is violated. To stop this linear growth, one possibility is that the observer should have access to the partner of Hawking mode (interior mode in Fig.  \ref{PenroseFSC}). In another words, a new region must be added to the entanglement wedge of radiation. According to the island proposal \cite{Penington:2019npb,Almheiri:2019psf}, the boundary of this new region is controlled by a new quantum extremal surface. To explore the existence of this quantum extremal surface, let us come back to the general expression for the finite part of the von Neumann entropy (\ref{Svn}) [or equivalently (\ref{SVN})], which at late times $t_a/\beta , t_b/\beta \gg 1$ reduces to
\bea
&&S_{\text{vN,fin}} = \frac{c}{6} \log\bigg[\frac{\beta^2 (r_0^2-a^2)}{4\pi^2 \hat{r}_+^2}\hspace{1mm}e^{-\frac{4\pi}{\beta}\left(t_a+t_b+b+r^{*}(a)\hspace{.05mm}\right)}
\cr\nonumber\\
&&\hspace{1cm}\times\left(e^{\frac{2\pi}{\beta}\left(t_a+t_b+2b\right)}-e^{\frac{2\pi}{\beta}\left(2t_a+b+r^{*}(a)\hspace{.05mm}\right)}-e^{\frac{2\pi}{\beta}\left(2t_b+b+r^{*}(a)\hspace{.05mm}\right)}+e^{\frac{2\pi}{\beta}\left(t_a+t_b+2r^{*}(a)\hspace{.05mm}\right)}\right)^{2}\bigg],
\eea
since the distance between the left wedge and the right wedge in Fig.  \ref{PenroseFSC} is very large and therefore 
\bea\label{OPE}
L(a_+,a_-) \simeq L(b_+,b_-) \simeq L(a_+,b_-) \simeq L(a_-,b+) \gg L(a_\pm,b_\pm). 
\eea 
Accordingly, in these times the generalized entanglement entropy becomes
\bea\label{SimpleSgentwoside}
&&S_{\text{gen}} = \frac{\pi \hspace{.5mm}a}{G_{\text{N,ren}}} + \frac{c}{6} \log\bigg[\frac{\beta^2  (r_0^2-a^2)}{4\pi^2 \hat{r}_+^2}\hspace{1mm}e^{-\frac{4\pi}{\beta}\left(t_a+t_b+b+r^{*}(a)\hspace{.05mm}\right)}
\cr\nonumber\\
&&\hspace{1.2cm}\times\left(e^{\frac{2\pi}{\beta}\left(t_a+t_b+2b\right)}-e^{\frac{2\pi}{\beta}\left(2t_a+b+r^{*}(a)\hspace{.05mm}\right)}-e^{\frac{2\pi}{\beta}\left(2t_b+b+r^{*}(a)\hspace{.05mm}\right)}+e^{\frac{2\pi}{\beta}\left(t_a+t_b+2r^{*}(a)\hspace{.05mm}\right)}\right)^{2}\bigg].
\eea
Now, from the extremization with respect to $a$ and $t_a$ one gets respectively,
\bea\label{twosidea}
&&\hspace{-1.5cm}\frac{\partial S_{\text{gen}}}{\partial a} = \frac{1}{3(r_0^2-a^2)}\bigg(\frac{3\pi(r_0^2-a^2)-c\hspace{.5mm}G_{\text{N,ren}}\hspace{1mm} a}{G_{\text{N,ren}}}+
\cr\nonumber\\
&&\hspace{-.5cm} + \frac{\hspace{1mm} c\hspace{1mm}e^{\frac{2\pi}{\beta}(t_a+t_b)}\left(e^{\frac{4\pi b}{\beta}}-e^{\frac{4\pi r^{*}(a)}{\beta}}\right)\hspace{1mm}a}{\left(e^{\frac{2\pi}{\beta}\left(t_a+t_b+2b\right)}-e^{\frac{2\pi}{\beta}\left(2t_a+b+r^{*}(a)\hspace{.05mm}\right)}-e^{\frac{2\pi}{\beta}\left(2t_b+b+r^{*}(a)\hspace{.05mm}\right)}+e^{\frac{2\pi}{\beta}\left(t_a+t_b+2r^{*}(a)\hspace{.05mm}\right)}\right)}\hspace{1.5mm}\bigg) = 0,
\eea
and
\bea\label{twosideta}
\hspace{.2cm}\frac{\partial S_{\text{gen}}}{\partial t_a} =-\frac{2\pi c}{3\beta}\frac{ \hspace{1mm}e^{\frac{2\pi}{\beta}\left(b+r^{*}(a)\right)}\hspace{1mm}\bigg(e^{\frac{4\pi t_a}{\beta}}-e^{\frac{4\pi t_b}{\beta}}\bigg)}{\left(e^{\frac{2\pi}{\beta}\left(t_a+t_b+2b\right)}-e^{\frac{2\pi}{\beta}\left(2t_a+b+r^{*}(a)\hspace{.05mm}\right)}-e^{\frac{2\pi}{\beta}\left(2t_b+b+r^{*}(a)\hspace{.05mm}\right)}+e^{\frac{2\pi}{\beta}\left(t_a+t_b+2r^{*}(a)\hspace{.05mm}\right)}\right)} =0.
\eea
The equation (\ref{twosidea}) implies that $a \neq r_0$. According to that and by noting to the expression of tortoise coordinate (\ref{tortoise}), we find that the only solution for Eq. (\ref{twosideta}) is $t_a = t_b$. For $t_a =t_b$, the generalized entanglement entropy (\ref{SimpleSgentwoside}) is more simplified to
\bea
S_{\text{gen}} = \frac{\pi \hspace{.5mm}a}{G_{\text{N,ren}}} + \frac{c}{6}\log\left[\frac{\beta^3 e^{-\frac{4\pi b}{\beta}}\left(e^{\frac{2\pi b}{\beta}}-\sqrt{ r_0^2-a^2}\right)^4}{8\pi^3 r_0}\right].
\eea
This allows an extremal surface at
\bea\label{QES}
a=r_0\left(1 - \frac{2c^2\hspace{.5mm}e^{-\frac{4\pi b}{\beta}}G_{\text{N,ren}}^2}{9\pi^2}\right),
\eea
where the generalized entanglement entropy (\ref{SimpleSgentwoside}) reads
\bea\label{Sfin}
S_{\text{gen}}= 2 S_{\text{th}} +\frac{c}{6}\log\left(\frac{\beta^3 e^{\frac{4\pi b}{\beta}}}{8\pi^3 r_0}\right)-\frac{2c^2 r_0 e^{-\frac{4\pi b}{\beta}}G_{\text{N,ren}}}{3\pi}+\mathcal{O}(G_{\text{N,ren}}^{2})
\eea
The first term is exactly twice the Bekenstein-Hawking entropy (\ref{temp and entropy of FSC}) and the other terms are the effect of quantum matter. In contrast to the late time result without the island (\ref{linear}), this is a constant that implies that the configuration with the island is preferred and the entropy stops growing. Interestingly, to resolve the information paradox we did not consider any gravitational backreaction\footnote{The energy (Hawking radiation) emitted from the FSC precisely balances the energy (radiation from the bath) falls in the FSC.} for the FSC, actually semiclassical gravity is sufficient. It is also worth noting that, according to (\ref{QES}), it looks surprising that the boundary of the island  is outside the horizon  because we can send a signal from the island to the bath. Since we are dealing with a coupled system (FSC+bath),  this coupling by time evolution mixes the degrees of freedom in bath and outside the FSC horizon and therefore there is no paradox in this case. But one might worry that, if we decouple the FSC from the bath, then we really encounter causality paradoxes. Indeed, to decouple the FSC from the bath one needs to a positive value of energy and by this energy flux into the FSC, the area of horizon becomes smaller\footnote{It can be seen by noting to the first law of thermodynamics for the FSC solution (\ref{first law CH}) together with expression for thermal entropy (\ref{temp and entropy of FSC}).} and the quantum extremal surface then lies behind the horizon, therefore again there is no paradox. Accordingly, even though the island is outside the horizon of the coupled system, it changes to the behind the horizon of the decoupled system. This is in agreement with arguments in \cite{Engelhardt:2014gca}. Based on generalized second law, it is argued in Ref. \cite{Engelhardt:2014gca} that the
quantum extremal surface should be behind the horizon.
\subsection{No island behind the horizon of coupled system}\label{behindH}
\begin{figure}[h]
\centering\includegraphics[scale=.6]{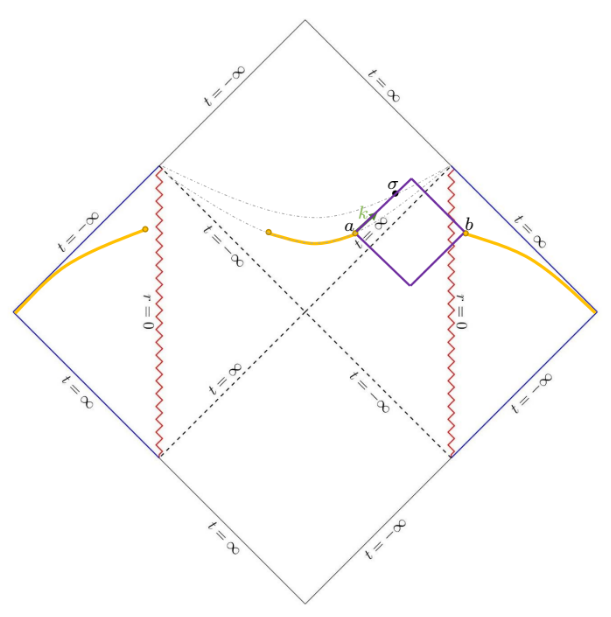} 
\caption{According to QFC, this setup is ruled out where $a$ is the assumed quantum extremal surface inside of the horizon and $\sigma$ denotes another codimesnion 2 surface. The $k$ is a null vector in the $\mathbf{v}$ direction and we take the region associated to QFC to be the causal domain of spatial interval between $a$ and $b$.  The boundary of the Causal domain (Wheeler-DeWitt patch) is shown in violet color. Since the right end point has been fixed to $b$, we cannot move the left end point beyond this boundary.}\label{QFC}
\end{figure}

In the previous subsection, we have assumed that the boundary of  island is located outside of the horizon, and we were able to confirm it by performing concrete calculations. In this subsection, we show that the island does not appear behind the horizon of the coupled system  (FSC+bath), i.e., in the top-wedge in Fig.  \ref{PenroseFSC}. By using the proper coordinates (\ref{UV}) for the top-wedge one gets
\bea\label{inside}
&&\hspace{-1cm}S_{\text{gen}} = \frac{\pi\hspace{1mm}a}{G_{\text{N,ren}}} + \frac{c}{6}\log\bigg[\frac{\beta^2 (a^2-r_0^2)}{4\pi^2\hat{r}_+^2}e^{-\frac{4\pi(t_a+t_b)}{\beta}}\left(-1+e^{\frac{4\pi t_a}{\beta}}\right)^2\left(1+e^{\frac{4\pi t_b}{\beta}}\right)^2
\cr\nonumber\\
&&\hspace{0.5cm}\times\hspace{1mm}\frac{\left(e^{\frac{2\pi}{\beta}\big(t_a+t_b+2b\big)}-e^{\frac{2\pi}{\beta}\big(2t_a+b+r^{*}(a)\big)}+e^{\frac{2\pi}{\beta}\big(2t_b+b+r^{*}(a)\big)}-e^{\frac{2\pi}{\beta}\big(t_a+t_b+2r^{*}(a)\big)}\right)^2}{\left(e^{\frac{2\pi}{\beta}\big(t_a+t_b+2b\big)}-e^{\frac{2\pi}{\beta}\big(b+r^{*}(a)\big)}+e^{\frac{2\pi}{\beta}\big(2t_a+2t_b+b+r^{*}(a)\big)}-e^{\frac{2\pi}{\beta}\big(t_a+t_b+2r^{*}(a)\big)}\right)^2}\hspace{.5mm}\bigg],
\eea
which at early times, $t_a/b ,t_b/b \ll 1$, is simplified to
\bea
S_{\text{gen}} = \frac{\pi\hspace{1mm}a}{G_{\text{ren}}} +\frac{c}{6} \log\left[\frac{\beta^2(a^2-r_0^2)}{4\pi^2 \hat{r}_+^2}\hspace{1mm}e^{-\frac{4\pi}{\beta}\big(t_a+t_b\big)}\left(-1+e^{\frac{4\pi  t_a}{\beta}}\right)^2\left(1+e^{\frac{4\pi t_b}{\beta}}\right)^2\right].
\eea 
By extremizing this early time general entropy with respect to the location of possible quantum extremal surface near the horizon, $a = r_0 +\epsilon$, we find the following equation
\bea
\frac{\pi}{G_{\text{ren}}} +\frac{c}{12 r_0}+\frac{c}{6\epsilon} = 0,
\eea 
which does not allow a consistent solution in the region $r_0 < a < \infty $. This implies that at early times the island is not generated for $r_0 < a < \infty $. Accordingly, in the following we check the existence of island region at late times. At late times, the general entropy (\ref{inside}) becomes
\bea\label{insidehorizon}
&&S_{\text{gen}} = \frac{\pi \hspace{.5mm}a}{G_{\text{N,ren}}} + \frac{c}{6} \log\bigg[\frac{\beta^2  (a^2-r_0^2)}{4\pi^2 \hat{r}_+^2}\hspace{1mm}e^{-\frac{4\pi}{\beta}\left(t_a+t_b+b+r^{*}(a)\hspace{.05mm}\right)}
\cr\nonumber\\
&&\hspace{1cm}\times\left(e^{\frac{2\pi}{\beta}\left(t_a+t_b+2b\right)}-e^{\frac{2\pi}{\beta}\left(2t_a+b+r^{*}(a)\hspace{.05mm}\right)}+e^{\frac{2\pi}{\beta}\left(2t_b+b+r^{*}(a)\hspace{.05mm}\right)}-e^{\frac{2\pi}{\beta}\left(t_a+t_b+2r^{*}(a)\hspace{.05mm}\right)}\right)^{2}\bigg],
\eea
where we have used again the relations (\ref{OPE}) to compute the bulk entropy. The extremization with respect to $a$ and $t_a$ gives, respectively,
\bea
&&\hspace{-1.2cm}\frac{\partial S_{\text{gen}}}{\partial a} = \frac{1}{3(a^2-r_0^2)}\bigg(\frac{3\pi(a^2-r_0^2)+c\hspace{.5mm}G_{\text{N,ren}}\hspace{1mm} a}{G_{\text{N,ren}}}+
\cr\nonumber\\
&&\hspace{-.5cm} - \frac{\hspace{1mm} c\hspace{1mm}e^{\frac{2\pi}{\beta}(t_a+t_b)}\left(e^{\frac{4\pi b}{\beta}}+e^{\frac{4\pi r^{*}(a)}{\beta}}\right)\hspace{1mm}a}{\left(e^{\frac{2\pi}{\beta}\left(t_a+t_b+2b\right)}-e^{\frac{2\pi}{\beta}\left(2t_a+b+r^{*}(a)\hspace{.05mm}\right)}+e^{\frac{2\pi}{\beta}\left(2t_b+b+r^{*}(a)\hspace{.05mm}\right)}-e^{\frac{2\pi}{\beta}\left(t_a+t_b+2r^{*}(a)\hspace{.05mm}\right)}\right)}\hspace{1.5mm}\bigg) = 0,
\cr\nonumber\\\nonumber\\
&&\hspace{-1.2cm}\frac{\partial S_{\text{gen}}}{\partial t_a} =-\frac{2\pi c}{3\beta}\frac{ \hspace{1mm}e^{\frac{2\pi}{\beta}\left(b+r^{*}(a)\right)}\hspace{1mm}\bigg(e^{\frac{4\pi t_a}{\beta}}+e^{\frac{4\pi t_b}{\beta}}\bigg)}{\left(e^{\frac{2\pi}{\beta}\left(t_a+t_b+2b\right)}-e^{\frac{2\pi}{\beta}\left(2t_a+b+r^{*}(a)\hspace{.05mm}\right)}+e^{\frac{2\pi}{\beta}\left(2t_b+b+r^{*}(a)\hspace{.05mm}\right)}-e^{\frac{2\pi}{\beta}\left(t_a+t_b+2r^{*}(a)\hspace{.05mm}\right)}\right)} =0.
\eea
It is clear that these two equations do not have any consistent solution for $t_a$ and $a$. Therefore, not only in early times but also at late times there is no island with boundary surfaces inside of the horizon, $r_0 < r < \infty $. This is consistent with the quantum focusing conjecture (QFC) \cite{Bousso:2015mna} as following. Let us assume that there is a quantum extremal surface behind the horizon. Since this is an extremum of generalized entanglement entropy, the first covariant derivative of $S_{\text{gen}}$ in the direction of null vector $k$ vanishes,
\bea\label{qes}
\nabla_{k}\hspace{.5mm}S_{\text{gen}}(\text{QES}) = 0.
\eea
According to QFC, the second covariant derivative of generalized entanglement entropy in the direction of null vector $k$ is not positive \cite{Bousso:2015mna},
\bea
\nabla^{2}_{k}\hspace{.5mm}S_{\text{gen}} \leq 0,
\eea
which together with (\ref{qes}) imply that by moving away from the QES in the direction of $k$ (right-moving light ray in Fig.  \ref{QFC}), the first derivative of generalized entanglement entropy must be less than zero. But away from the QES, for $S_{\text{gen}}$ in (\ref{insidehorizon})  one can see that
\bea\label{ResultQFC}
\nabla_{k}\hspace{.5mm}S_{\text{gen}} = \frac{2\pi^2}{\beta G_{\text{N,ren}}}\frac{r^2-r_0^2}{r}+\mathcal{O}(G^{0}_{\text{N,ren}}) > 0.
\eea
This contradiction stems from the fact that we wrongly assumed that there is a QES behind the horizon. On the other hand, if we assume that the QES is outside of the horizon then by using (\ref{SimpleSgentwoside}) we can find the first derivative of $S_{\text{gen}}$ in the right-moving light ray direction that it again becomes (\ref{ResultQFC}) and that is clearly negative for $r < r_0$, consistent with QFC.
\subsection{Page time and scrambling time}
\begin{figure}[h]
\centering\includegraphics[scale=.6]{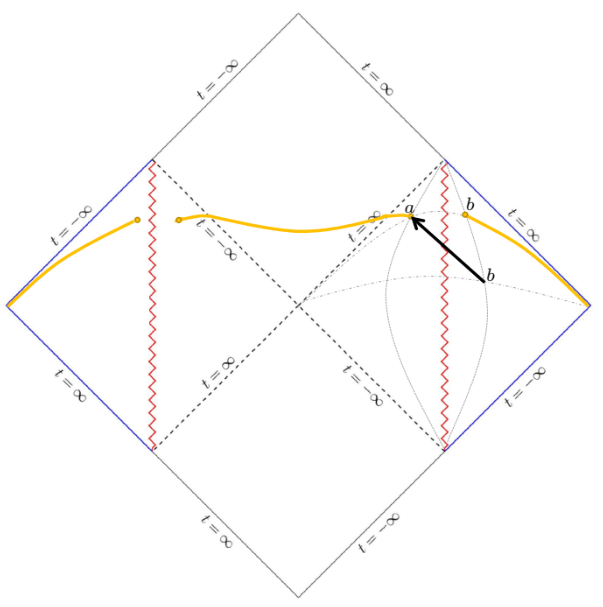} 
\caption{A signal that is thrown into the FSC comes up in the radiation degree of freedom after scrambling time.}\label{Scrambling}
\end{figure}

By equating the entropy (\ref{Sgentwosidewithout}) with the generalized entropy (\ref{Sfin}), one can determine the Page time which is a time when the entropy stops growing. Doing so, at late times $t_b/\beta \gg 1$ , we arrive at
\bea
t_{\text{Page}} = \frac{3\beta }{\pi c}S_{\text{th}}+\frac{\beta}{4\pi}\log\left(\frac{\beta\hspace{1mm}e^{\frac{4\pi b}{\beta}}}{2\pi r_0}\right)-\frac{c\beta r_0 \hspace{1mm}e^{-\frac{4\pi b}{\beta}}G_{\text{N,ren}}}{\pi^2}+\mathcal{O}(G^{2}_{\text{N,ren}})
\eea
Moreover, according to (\ref{QES}), we can determine the scrambling time. The scrambling time is defined based on the minimum time required to retrieve the information after sending the information into the black hole. Since in our setup the radiation degrees of freedom are encoded in the union of $R\cup I$, the signal thrown into the FSC comes up in the radiation degree of freedom after signal reaches to the island. Let us assume that the observer is sitting at the radius $b$ and at time $t_0$ sends a signal into the FSC, see Fig.  \ref{Scrambling}. This signal will get to the island with a boundary at radius $a$ and time $t_a$. The distance between these two points in the ingoing null direction is given by
\bea
\mathbf{v}(t_0,b) - \mathbf{v}(t_a,a) = \big(t_0+r^{*}(b)\big) - \big(t_a+r^{*}(a)\big),
\eea
therefore the time difference between the boundary of the island and the initial time is given by
\bea
t_a -t_0 = \big(r^{*}(b)-r^{*}(a)\big)-\big(\mathbf{v}(t_0,b) - \mathbf{v}(t_a,a)\big).
\eea
Since the $\mathbf{v}(t_a,a)$ should be equal or greater than $\mathbf{v}(t_0,b)$ to have signal in island,  therefore the minimum time to retrieve the information is given by
\bea
t_{\text{scr}} \equiv t_a -t_0 = r^{*}(b)-r^{*}(a),
\eea
which according to (\ref{tortoise}) and (\ref{QES})  becomes
\bea
t_{\text{scr}} = \frac{\beta}{2\pi}\log S_{\text{th}}+ 2 b -\frac{\beta}{2\pi}\log\left(\frac{c \hspace{.5mm}r_0^2}{3}\right) \simeq  \frac{\beta}{2\pi}\log S_{\text{th}}
\eea
The leading dependence is universal \cite{Shenker:2013pqa} and this $t_{\text{scr}}$ actually is the fast scrambling time. This scrambling time can also be written as 
\bea
t_{\text{scr}} = \frac{\beta}{2\pi}\log \left(\frac{3}{G_{\text{N,ren}}}\right) + 2b +\frac{\beta}{2\pi}\log\left(\frac{\pi}{2c\hspace{.5mm} r_0}\right) \simeq \frac{\beta}{2\pi}\log c_{M},
\eea
which is in agreement with \cite{Bagchi:2021qfe} and $c_M = 3/G_{\text{N,ren}}$ is the central charge of (Bondi-Metzner-Sachs)$_3$ algebra associated to the asymptotic symmetry group at null infinity \cite{Barnich:2006av}. The study in \cite{Bagchi:2021qfe} is based on evaluating  special four-point out-of-time-ordered correlation functions relevant to the (quantum) chaos and scrambling time is defined as the timescale
where the global block expansions fail. 
\subsection{Extracting information from the island}\label{Teleportation}
In Sec. 3.1 \ref{islandoutside}, we observed that to resolve the information paradox in our setup, we need to include a new region (island) to the entanglement wedge of Hawking radiation.  One might expect that it might be possible to also extract the information in the island. Even thought extracting information in general could be very difficult, but in our setup there is a simple way to do it following \cite{Gao:2016bin}. As we have mentioned previously, the state of the whole system (FSC+bath) is a thermofield double state with the following standard representation,
\bea\label{TFD}
\mid\hspace{-1mm}\text{TFD}(t_L,t_R)\big{\rangle} = \frac{1}{\sqrt{Z(\beta)}}\hspace{1mm} e^{-i(\hat{H}_L t_L+\hat{H}_R t_R)}  \sum_{n} e^{-\beta E_n/2} \mid \hspace{-1mm}E_n\big{\rangle}_{L}\mid \hspace{-1mm}E_n\big{\rangle}_{R}.
\eea
This state has a large bipartite entanglement  between the right and left side since the von Neumann entropy of the reduced density matrix $\rho_{R} = \text{Tr}(\rho_{\text{TFD}})_{L}$ is given by
\bea
S_{\text{vN}}(\rho_{R}) = S_{\text{th}}  \sim 1/G_{\text{N}}.
\eea
This large amount of entanglement creates a connected geometry, Einstein-Rosen bridge. Even in presence of this wormhole between the two sides, a signal from the left-side island cannot be seen in the right side, according to the causality. The absence of this communication stems also from the absence of interaction between boundary QFTs. Now let us turn on an interaction between the boundary QFT in the right side and boundary QFT in the left side at the time $t=0$, by adding the operator $\exp{(i g \hat{X}_{L}(0) \hat{X}_{R}(0))}$ to the total Hamiltionian in (\ref{TFD}). According to the gauge/gravity duality, this means that in the gravity we have turned on a nonlocal interaction $\exp{(i g \Phi_{L}(0) \Phi_{R}(0))}$, where the fields $\Phi$ are sources for the operators $\hat{X}$. In the simplest case, where we have one free scalar field and in the limit $G_{N}\rightarrow 0$, the energy-momentum tensor of scalar theory after turning on this coupling becomes
\bea\label{Tuu}
\big\langle T_{\mathbf{u}\mathbf{u}}\big\rangle_{g}  = \big\langle e^{-i g \Phi_{L}(0) \Phi_{R}(0)}\hspace{1mm} T_{\mathbf{u}\mathbf{u}}\hspace{1mm} e^{i g \Phi_{L}(0) \Phi_{R}(0)}\big\rangle_{g=0}
\eea
It is clear that by adjucting the coupling constant $g$ the deformed energy can become negative. It can be seen by noting that (\ref{Tuu}) at first order of perturbation becomes
\bea
\big\langle T_{\mathbf{u}\mathbf{u}}\big\rangle_{g}  = -i g\hspace{.5mm} \big\langle\hspace{.2mm}\big[\Phi_{L}\Phi_{R}\hspace{.5mm},\hspace{.5mm} \partial_{\mathbf{u}}\Phi\partial_{\mathbf{u}}\Phi\big]_{g=0}\hspace{.2mm}\big\rangle = -ig \left[\Phi_{L}\hspace{.5mm},\hspace{.5mm}\partial_{\mathbf{u}}\Phi\right]_{g=0} \big\langle \Phi_{R}\hspace{1mm}\partial_{\mathbf{u}}\Phi\big\rangle_{g=0}+ L \leftrightarrow R
\eea 
The commutator gives the delta function (shockwave in the $\mathbf{v}$ direction) in the location of the field and another term is related to the causal propagator. It is worthwhile to emphasize that while the initial deformations are localized in the baths, the coupled Hamiltonian in this protocol is crucial for propagating the negative energy into the bulk. Albeit this shock wave has negative energy but it increases the size of horizon when entering the FSC, since we have a nonstandard first law of thermodynamics for the FSC solution,the minus sign in  (\ref{first law CH}). This increase in the size of the horizon is given by\footnote{The first relation is actually the first law (\ref{first law CH}).}
\bea\label{DeltaV}
\Delta \mathbf{V} \sim
-G_{\text{N,ren}} \int T_{\mathbf{u}\mathbf{u},g}\hspace{1mm}d\mathbf{u} =- g\hspace{.5mm} \mathcal{G}_{\text{shock}}\hspace{.5mm} G_{\text{N,ren}}+\mathcal{O}(G_{\text{N,ren}}^2),
\eea
where $\mathcal{G}$ is related to causal propagator. Moreover, the distance between the quantum extremal surface $a$, (\ref{QES}), and the past horizon is given by
\bea
\Delta \mathbf{V} = \mathbf{V}(t_a,a) -\mathbf{V}(-\infty,r_0) = \frac{2c\hspace{.5mm} r_0}{3\pi} e^{-\frac{2\pi}{\beta}(b-t_b)}\hspace{.5mm}G_{\text{N,ren}}+\mathcal{O}(G_{\text{N,ren}}^2),
\eea  
which implies that by choosing the $g$ properly,  the nonlocal interaction can produce enough negative energy to pull the island into the causal contact with the left system. Since $\Delta \mathbf{V}$ (\ref{DeltaV}) is of order  $G_{\text{N,ren}}$, the wormhole becomes slightly traversable\footnote{The way we glue the two boundaries breaks the time killimg symmetry in the bulk and the signal cannot back to a time. Therefore, it exists no closed time-like curves in our setup.}. Accordingly, the information stranded in the right-side island can be rescued and detected in the left bath. This protocol will be successful if we also make sure that the information in island is transferred correctly to the other side, i.e.,
we have a procedure which for any $\mid\hspace{-1mm}\Psi \rangle$ implements  
\bea\label{QTT}
\big{|}\hspace{-.1mm}\Psi \big\rangle_{I}\hspace{1mm} \big{|}\hspace{-.1mm}0\big\rangle_{\text{Rad}_{L}} \hspace{.5cm}\longrightarrow\hspace{.5cm} \big{|}\hspace{-.1mm}0\big\rangle_{I}\hspace{1mm} \big{|}\hspace{-.1mm}\Psi\big\rangle_{\text{Rad}_{L}} 
\eea
or more generally
\bea
\big{|}\hspace{-.1mm}\Psi \big\rangle_{I} \hspace{1mm}\big{|}\hspace{-.1mm}0\big\rangle_{\text{Rad}_{L}} \hspace{.5cm}\longrightarrow \hspace{.5cm} \big{|}\hspace{-.1mm}0\big\rangle_{I} \hspace{1mm}U_{\text{Rad}_{L}} \big{|}\hspace{-.1mm}\Psi\big\rangle_{\text{Rad}_{L}} 
\eea
where $U$ is a unitary operator\footnote{The receiver can get back to the evolution (\ref{QTT}) by action with $U^{\dagger}_{\text{Rad}_{L}}$.}. The way to test it is
introducing an additional auxiliary system $A$, of the same dimensionality as island $I$ in the right side and radiation in the left bath,
and maximally entangle it with $I$. According to the linearity, we then have the evolution 
\bea
\frac{1}{\sqrt{\mid I\mid}} \sum_{i} \big{|}\hspace{-.1mm}i \big\rangle_{I}\hspace{1mm}\big{|}\hspace{-.1mm}0 \big\rangle_{\text{Rad}_{L}}\big{|}\hspace{-.1mm}i \big\rangle_{A} \hspace{.5cm}\longrightarrow\hspace{.5cm} \big{|}\hspace{-.1mm}0 \big\rangle_{I}\hspace{1mm}\frac{1}{\sqrt{\mid I\mid}} \sum_{i} U_{\text{Rad}_{L}}\big{|}\hspace{-.1mm}i \rangle_{\text{Rad}_{L}}\big{|}\hspace{-.1mm}i \rangle_{A},
\eea
which transfers the purification of $A$ from $I$ to $\text{Rad}_{L}$. Now, if 
the final states $\rho_{IA}$ will be close to  $\rho_I \otimes \rho_A$ in the trace norm, we can claim that the transfer is successful \cite{Harlow:2014yka}. Since all we have done is valid in the low-energy limit (semiclassical regime) and we do not know the microstates of the island, therefore we can not check this test concretely at the moment. 
Last but not least, it is also worth noting that with this protocol we cannot transfer the information forever since, for that purpose, we  not only  need to create large amount of negative energy but also need many signals from the island. As both of them can backreact on the geometry, accordingly we lose our control on the simple background spacetime.
\begin{figure}[h]
\centering\includegraphics[scale=.6]{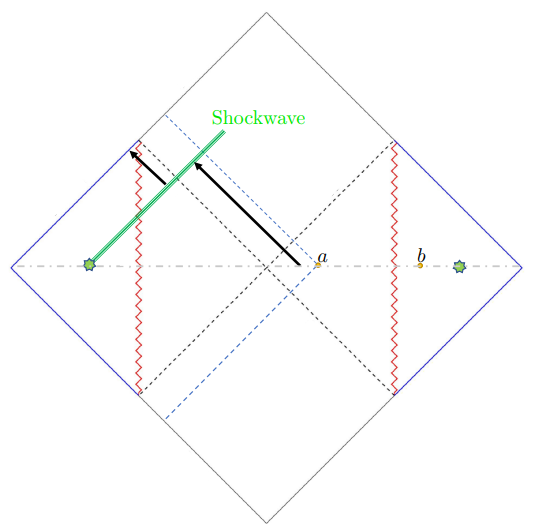} 
\vspace{-.3cm}
\caption{Recovering the information of island through a quantum teleportation protocol. By choosing the proper sign for the coupling $g$, one can create two shock waves with negative energy in the bulk where one of them is presented here. After changing the location of horizon, the signal from the right-side island by reaching to the negative energy shockwave gets time advances and comes out on the other side.  
}\label{QT}
\end{figure} 
\section{Discussion}\label{Discussion}
In this paper we study the eternal two-sided FSC solution that is coupled to a nongravitational bath. The whole system starts in a pure state and then evolves in time. Conceptually, it is similar to a collapse of matter prepared in the pure state to create a black hole and then its evaporation. But in comparison to the real case, our setup is simpler since the background solution is simpler to find and therefore we can present a version of information paradox where the quantum state and geometry are very simple. Actually, without encountering the conceptual problems related to the backreaction of matter fields on the geometry, we can also resolve this paradox by finding new regions (islands) in the gravitational system. It is worthwhile to emphasize again that the whole procedure is based on semiclassical approximation without any need to the information about the quantum UV completion. Furthermore, apart from a configuration with a single island, it might also be configurations with more islands, though in this paper we did not consider them. At late times, these extra configurations would not have a dominant contribution since a configuration with a single island already agrees with twice that of the Bekenstein-Hawking entropy of the eternal two-sided cosmological solution. But around the Page time, they might contribute and accordingly the sharp change of general entropy may be smoothed away.\\

More importantly, the minus sign in the first law of thermodynamics for the FSC solution, (\ref{first law CH}), is a source for two interesting observations: (i) in our setup, the island is located outside of the horizon of coupled system (FSC+bath). When we decouple the FSC from the bath, two positive energy shock waves are created that, by entering the cosmological horizon, decrease the size of the horizon. Accordingly, the island changes to behind the horizon in agreement with the general argument of Engelhardt and Wall. The decreasing size of the horizon by absorbing the positive energy is related to this minus sign. (ii) To extract information from the island, we need a protocol by which the wormhole becomes traversable. This can be provided by creating the negative energy shockwave in the bulk. This negative energy shockwave increases the size of horizon and after that information can be causally transferred from the island to the bath. Again, increasing the size of horizon by absorbing the negative energy is related to that minus sign. More precisely, both of the above observations are in agreement with quantum focusing conjecture and therefore this minus sign can also affect the (averaged) null energy condition for the FSC geometry which is worthwhile, to be explored further.\\

Furthermore, the Ryu-Takayanagi formula and its extension in presence of quantum matter can be obtained by replica method for the gravitational path integral \cite{Lewkowycz:2013nqa,Faulkner:2013ana}. This means that the island formula is also calculable by using the replica method  \cite{Penington:2019kki,Almheiri:2019qdq}. The later one implies that there are geometries connecting the different replicas which are know as replica wormholes. These geometries are used in the JT gravity to analyze the late time behavior of correlation functions \cite{Saad:2019pqd} and spectral form factor  \cite{Saad:2018bqo,Saad:2019lba}. Intriguingly, these wormholes can give a small overlap (of order $e^{-S_{\text{th}}}$) between na\"ively orthogonal bulk states and this small correction to the Hawking radiation can restore the unitarity in evaporation process. But apart from these very fascinating characteristics, they lead to a factorization puzzle. Since, there is no interaction term between the dual QFTs in the two sides therefore the partition function 
of combined system is actually a product of the partition function of left and right systems, $Z_{LR} = Z_L Z_R$. However, it seems that the presence of replica wormholes in the bulk implies that $Z_{LR} \neq Z_L Z_R$. A resolution suggested was that in presence of wormholes the bulk theory is dual to an ensemble of field theories \cite{Saad:2019lba}. For the FSC case, we have the same factorization problem and it might be the same resolution as for the JT gravity, which is interesting to be explored. More importantly, one can ask what happens to the wormholes connecting the decoupled system when we focus on just one element of the ensemble. For the dual of JT theory, which is the SYK model, it was shown that not only those wormhole saddles persist but also new  saddles exist, which are named as half-wormholes \cite{Saad:2021rcu}. Exploring this new saddles for the FSC case would be also very interesting.\\

Last but not least, in Sec. 3.4 \ref{Teleportation} we have implicitly assumed that there are local gauge invariant excitations in the island region and we want to extract information about them. In the gravitational system (FSC without bath) as a gauge theory, in order to define a gauge invariant operator one needs a dressing procedure. Accordingly, in order to define a gauge invariant operator for the island $I$, even a spatial geodesic should pass through the complement region $\Sigma-(R\cup I)$ to reach the radiation region $R$, where $\Sigma$ here denotes the overall Cauchy slice. This implies that to construct this gauge invariant operator, we not only need the information of the entanglement wedge of radiation ($R\cup I$) but also the information of the entanglement wedge of its complement. But this is in contradiction with the known principle \cite{Raju:2020smc} that the algebra of an entanglement wedge should be closed and commute with the algebra
of its complement. According to Refs. \cite{Geng:2021hlu,Geng:2020qvw}, the source of this puzzle seems to be whether or not there is massless graviton in the setup. If there is massless graviton, then we really encounter the problem since in the procedure of dressing we connect the entanglement wedge of radiation to its complement. But, if there is no  massless graviton, then the necessary Green function to define the dressing is a decaying function\footnote{Some criticisms on this issue can be found in \cite{Krishnan:2020oun,Krishnan:2020fer,Ghosh:2021axl}.}. Hence it might not be any connection between the entanglement wedge of radiation with its complement and consequently there is no puzzle\footnote{More precisely, in presence of a mass term, the equation that describes the  linearized graviton $h_{ij}$ on   $\bar{g}_{ij}$ background together with the energy density $\rho$ of excitations, $\mathcal{F}(\bar{g}_{ij},h_{ij}) +m^2 h_{ii} \sim G_{N}\hspace{.2mm}\rho$, has not a gradient form. Therefore, the integral of energy density over a volume cannot be expressed as a boundary term.}. Interestingly, in the similar setup, AdS spacetimes in $d>2$ dimensions, the graviton picks up mass in coupling to the nongravitational bath \cite{Porrati:2001db,Aharony:2003qf,Porrati:2003sa}. The reason is that the energy-momentum tensor of gravitational system on the boundary of AdS is no longer conserved. It is worth noting that if there is for example an $U(1)$ charged excitation in the island, there is no problem to associate a gauge invariant operator to it since for such gauge theory we have negative and positive charges together. In a gravitational system, there is just one charge with a fixed sign. Our setup is similar to the AdS case, where we couple FSC solution in $d=3$ dimensions\footnote{In $d=3$ dimensions, there is a notion of
“graviton.”} to the bath. By this coupling and allowing the modes to travel freely to the bath, the energy-momentum tensor in the gravitational region (FSC) is no more conserved and the graviton can becomes massive, accordingly there is no puzzle also in our setup. Of course, checking this guess more accurately needs concrete calculations such as the one for the AdS and we hope to address it in our future works. 

\subsubsection*{Acknowledgements}
The authors would like to thank Mohsen Alishahiha,
Kuroush Allameh, Amin Faraji Astaneh, Mostafa Ghasemi, S. Sedigheh Hashemi, Zahra Kabiri, Reza Pirmoradian, Suvrat Raju, Behrad Taghavi, and Amir Hossein Tajdini for useful comments and fruitful discussions. The work of A.N. was supported by Iran National Science Foundation (INSF) Grant No. 98014192. The work of S. A. and R. F. is based upon research funded by Iran National Science Foundation (INSF) under Project No 4003108. Finally we would like to thank the referee for her/his constructive comments.
\appendix

\end{document}